\documentclass[sigchi-a, authorversion]{acmart}
\usepackage{booktabs} 
\usepackage{ccicons}  
\usepackage{caption}
\usepackage{subcaption}

\setcopyright{none}

\acmDOI{}

\acmConference[FIS'18]{Fachpraktikum Interaktive Systeme}{Summer 2018}{University of Stuttgart, Germany}
\acmYear{2018}
\copyrightyear{2018}
\acmPrice{00.00}

\usepackage{multicol}
\usepackage{balance}    

\begin{document}
\title{Gesture Recognition based on Long-Short Term Memory Cells (LSTM) using Smartphone IMUs}

\author{Yuvaraj Govindarajulu}
\affiliation{%
  \institution{University of Stuttgart}
  \city{Stuttgart}
  \country{Germany} }
\email{st158830@stud.uni-stuttgart.de}

\author{Raja Rajeshwari Raj Kumar}
\affiliation{%
  \institution{University of Stuttgart}
  \city{Stuttgart}
  \country{Germany} }
\email{st159607@stud.uni-stuttgart.de}


\renewcommand{\shortauthors}{F. Author et al.}

%
%

\begin{abstract}
Over the last few decades, Smartphone technology has seen significant improvements. Enhancements specific to built-in Inertial Measurement Units (IMUs) and other dedicated sensors of the smartphones(which are often available as default) such as- Accelerometer, Gyroscope, Magnetometer, Fingerprint reader, Proximity and Ambient light sensors have made devices smarter and the interaction seamless. Gesture recognition using these smart phones have been experimented with many techniques.
In this solution, a Recurrent Neural Network (RNN) approach, LSTM (Long-Short Term Memory Cells) has been used to classify ten different gestures based on data from Accelerometer and Gyroscope. 
Selection of sensor data (Accelerometer and Gyroscope) was based on the ones that provided maximum information regarding the movement and orientation of the phone. Various models were experimented in this project, the results of which are presented in the later sections. Furthermore, the properties and characteristics of the collected data were studied and a set of improvements have been suggested in the future work section.

\end{abstract}


\maketitle

\section{Introduction \& Related Work}
The Human-Computer Interaction community has made significant progress in the last few years, enhancing the use of smartphones through natural ways of interaction. Among other techniques, research on gesture based interactions using smartphones, smart watches and camera based systems have shown an appealing as well as successful alternative to the conventional contact-based interaction.

The Smartphone industry has exhibited timely growth in incorporating necessary Hardware sensors and signal processing units without altering the manufacturing cost and selling price, thereby making it viable to people around the globe. In this context, IMUs have become a default unit in the smartphone these days. Also, these systems have proven to be robust which makes it possible to readily integrate gesture recognition on unmodified smartphones.

A number of publications describe various devices such as a pen \cite{JingYang2004} and an application specific hardware unit\cite{Xu2012} incorporating necessary IMUs . Articles such as \cite{K2015}, \cite{JongKOh} and \cite{BLSTM_HAR} show the benefits of using Gyroscope data together with Accelerometer data for motion based recognition on Smartphones. The measured signals from sensors such as Accelerometer and Gyroscope (used in this project) implicitly encode the gesture description. Lack of uni-vocal link between the gesture motion and the measured quantities calls for the use of Machine Learning (ML) approaches for recognition and classification.\cite{Belgioioso2014}. The paper by G. Belgioioso\cite{Belgioioso2014} compares the various ML methods involving feature extraction and their performances. The focus of this project however is to explicitly make use of the IMUs in the Smartphone, which was majorly inspired by the results from the above mentioned articles.

\begin{marginfigure}
	\includegraphics[width=\marginparwidth]{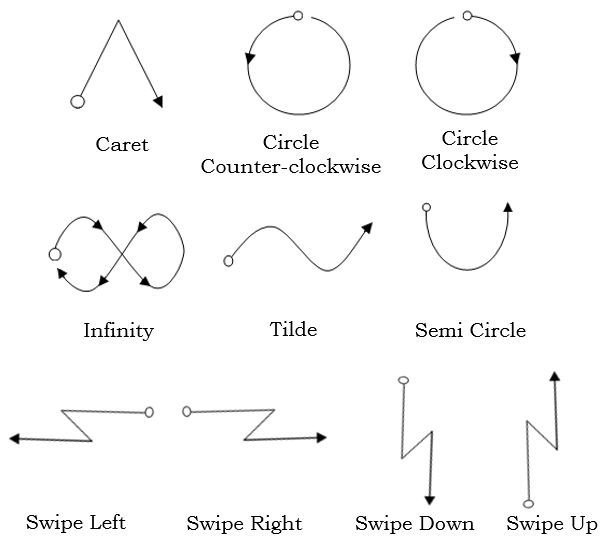}
	\caption{Set of gestures considered for this project.}
	\label{fig:gestures02}
\end{marginfigure}

Furthermore, Deep learning for Gesture Recognition has proven to be successful through its data-driven approach for learning efficient discriminative features from raw data, which enables feature abstraction. It avoids reliance on heuristics and hand-picked features together with better scaling for multi-dimensional motion recognition. Long-Short Term Memory (LSTM) cells, a variant of RNNs, combat the vanishing gradient issues and are good at handling time-series problems. It is shown that LSTM cells have a particular structure for remembering longer information.\cite{BLSTM_HAR} 

Previous works have shown the use of various methods for Gesture recognition. In this project, a sliding window approach on sensor time-series data of different lengths is used with an LSTM Model. The sliding window was moved through the time-series to cut it into samples of equal length. The 3-axes of the accelerometer and gyroscope respectively were fed as a 6-Dimensional input to the model. The classification involved a total of 10 different gestures(Fig:\ref{fig:gestures02}). The gestures were carefully selected such that they were not similar to the movements made when the phone is held by the user during normal operations. The properties of the data collected were analyzed during the course of the project to unveil further useful implications.

\begin{marginfigure}
	\includegraphics[width=\marginparwidth, scale=0.6]{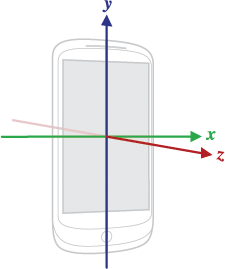}
	\caption{ Coordinate system (relative to a device) that's used by the Sensor API.\cite{sensorsoverview}}
	\label{fig:axis_device}
\end{marginfigure}

\section{Data Collection Study}
During the data collection phase, smartphone Accelerometer and Gyroscope sensor data were recorded as a time-series in csv file format for each gesture performed by the user. An Android application was developed that recorded the sensor data based on a touch event. One of the challenges was to correctly extract the important parts of the recorded data. Another device(Phone-2) that was operated by the supervisor was used to mark the begin and end time-stamps of the recorded gestures, in order to avoid additional noise that would be captured if the user were to start and stop the gestures himself.

\begin{marginfigure}
	\includegraphics[width=\marginparwidth,scale=0.85]{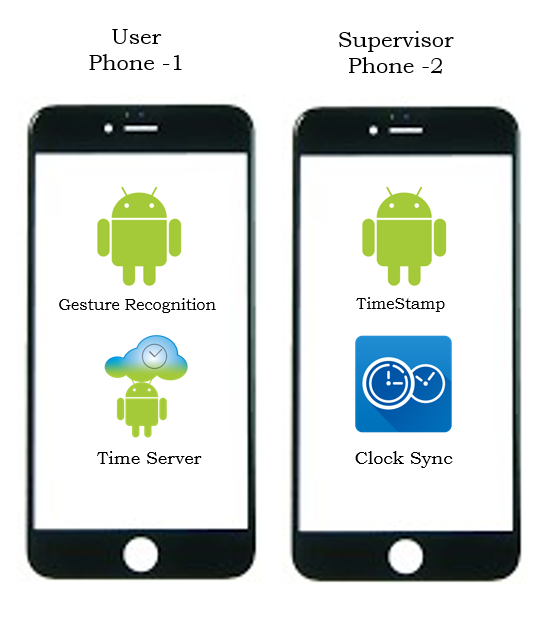}
	\caption{Representation of the Applications running on the phones.}
	\label{fig:Apparatus}
\end{marginfigure}

\subsection{Apparatus}
Two Smartphones (identical) Nexus5X\cite{Wiki:Nexus5X} were used. Both the phones run on the same Android versions - Oreo 8.1.0 API Level 27. 
Gesture Training and TimeStamp applications were installed on Phone-1 and Phone-2 respectively. For Clock Synchronization between the phones, an NTP Server runs on Phone-1 using TimeServer Application\cite{TimeServer}. NTP Server address was then fed into the ClockSync Application\cite{ClockSync} on Phone-2.(Fig:\ref{fig:Apparatus})

Recording on the Gesture training and the Timestamp applications were started simultaneously by the user and the supervisor. Gesture Training Application was used to collect user information, generate a random pattern of gestures one after another and record the sensor data for the complete gesture set (set of 10 gestures). Upon every sensor value change, the application records current system timestamp together with 3-axes Accelerometer and 3-axes Gyroscope values in a CSV file. The timestamps of the begin and end of each gesture were captured in another csv file by the TimeStamp application. Finally, the two csv files were used for preprocessing.

\begin{figure}
	\begin{subfigure}[b]{0.24\textwidth}
		\includegraphics[width=\textwidth]{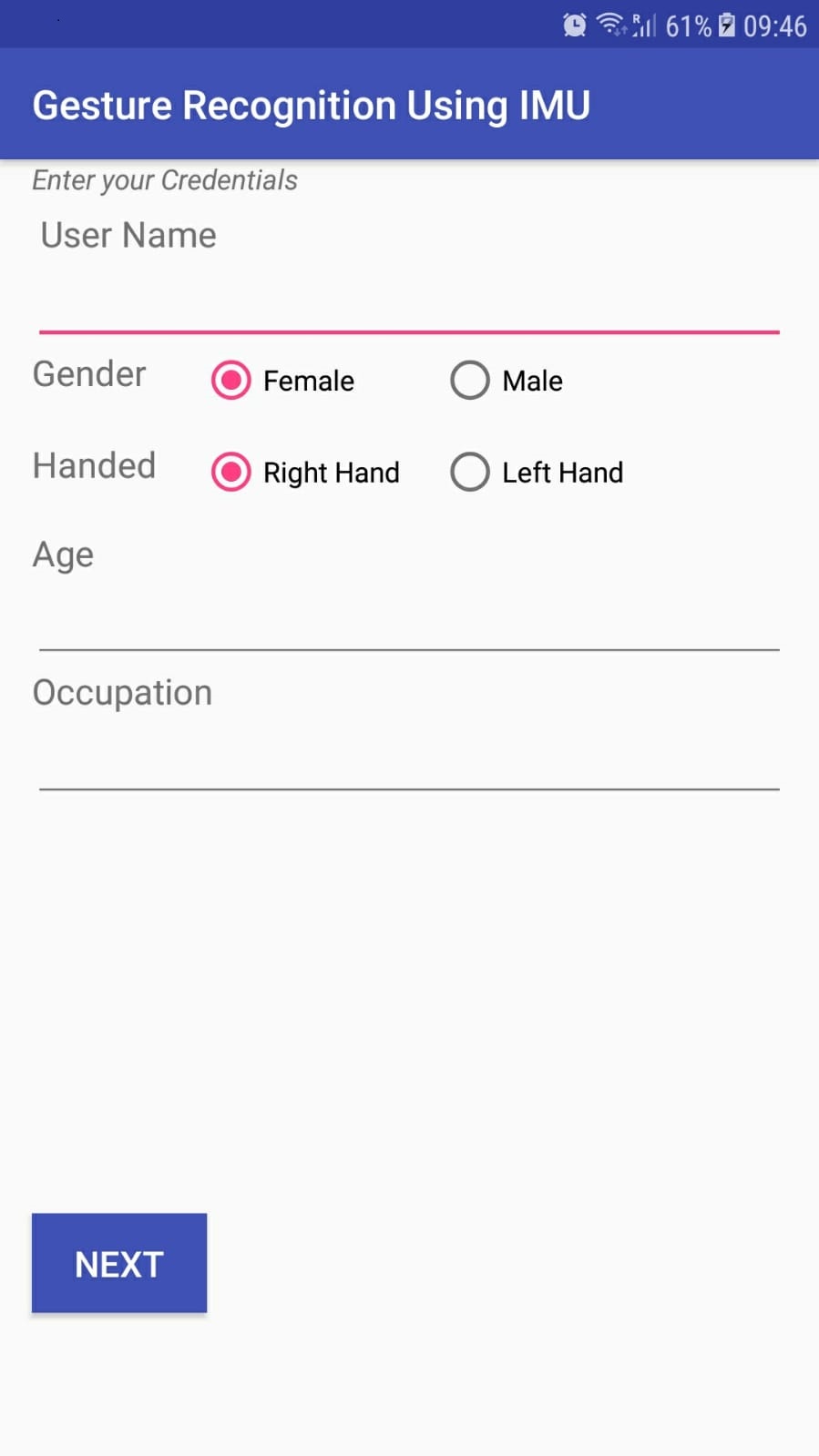}
		\caption{User-data collection}
		\label{fig:UserInfo}
	\end{subfigure}
	\begin{subfigure}[b]{0.24\textwidth}
		\includegraphics[width=\textwidth]{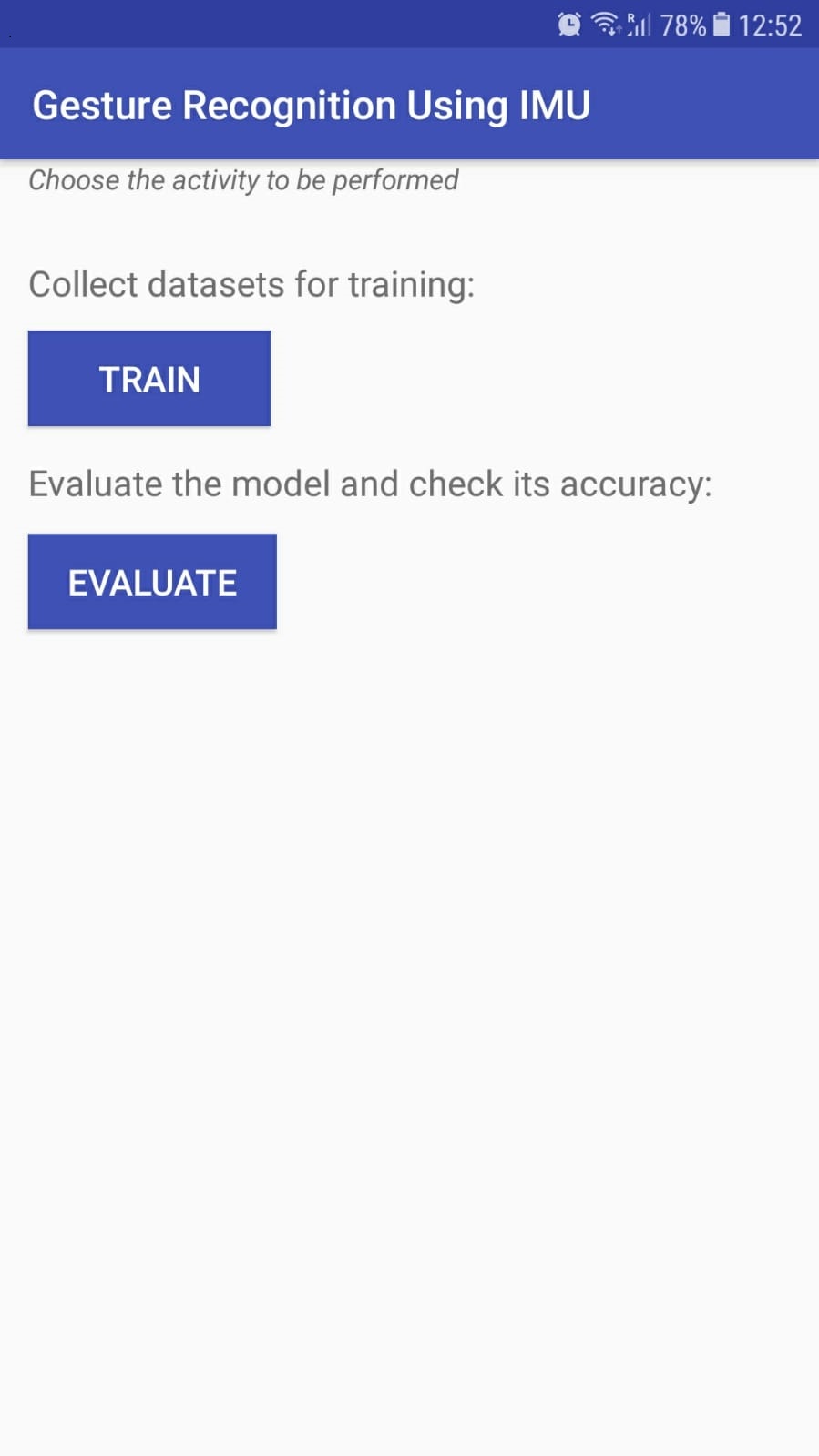}
		\caption{Selection Page}
		\label{fig:MainPage}
	\end{subfigure}
	\begin{subfigure}[b]{0.24\textwidth}
		\includegraphics[width=\textwidth]{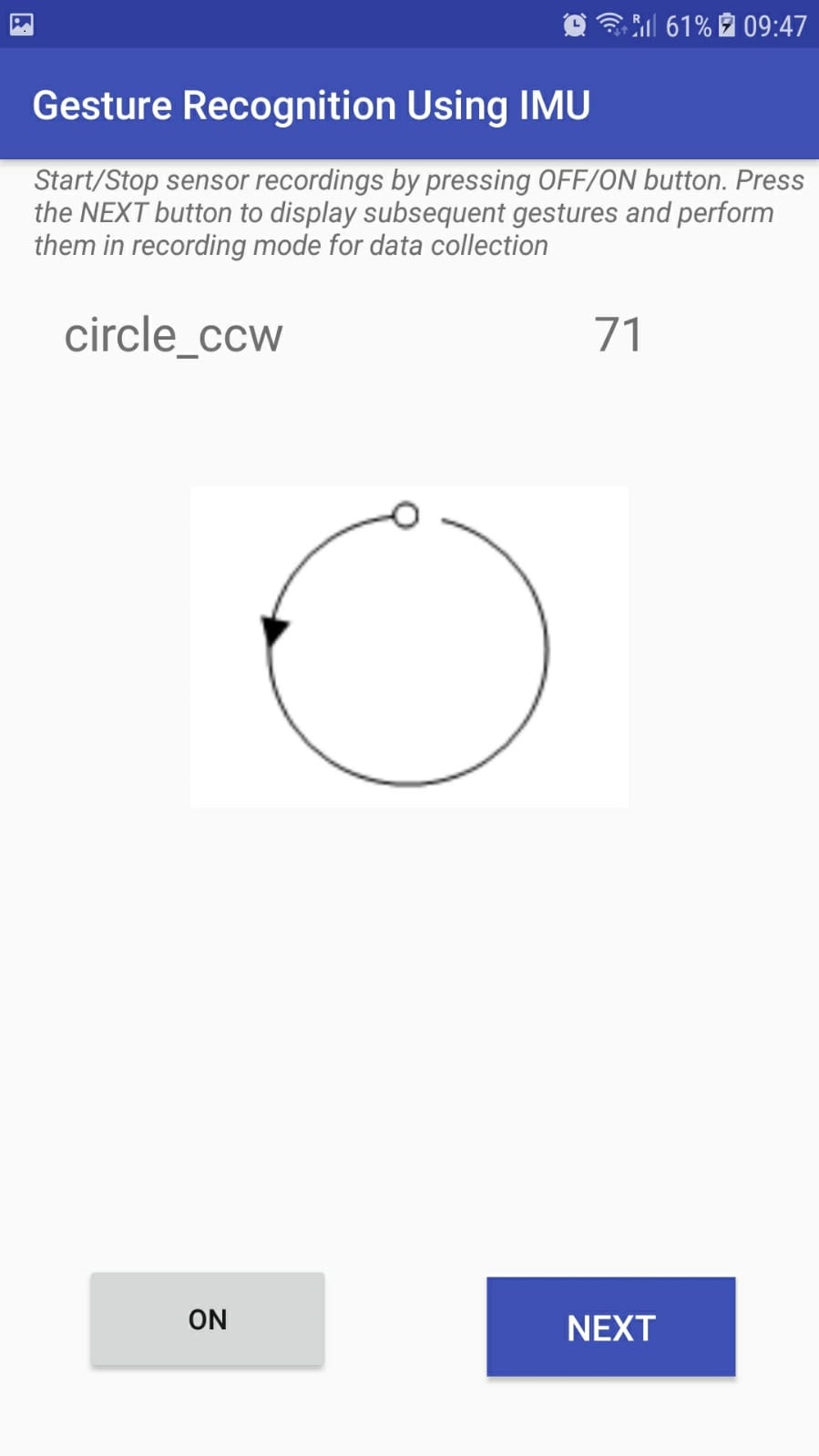}
		\caption{Training Page}
		\label{fig:GestureTrain}
	\end{subfigure}
	\begin{subfigure}[b]{0.24\textwidth}
		\includegraphics[width=\textwidth]{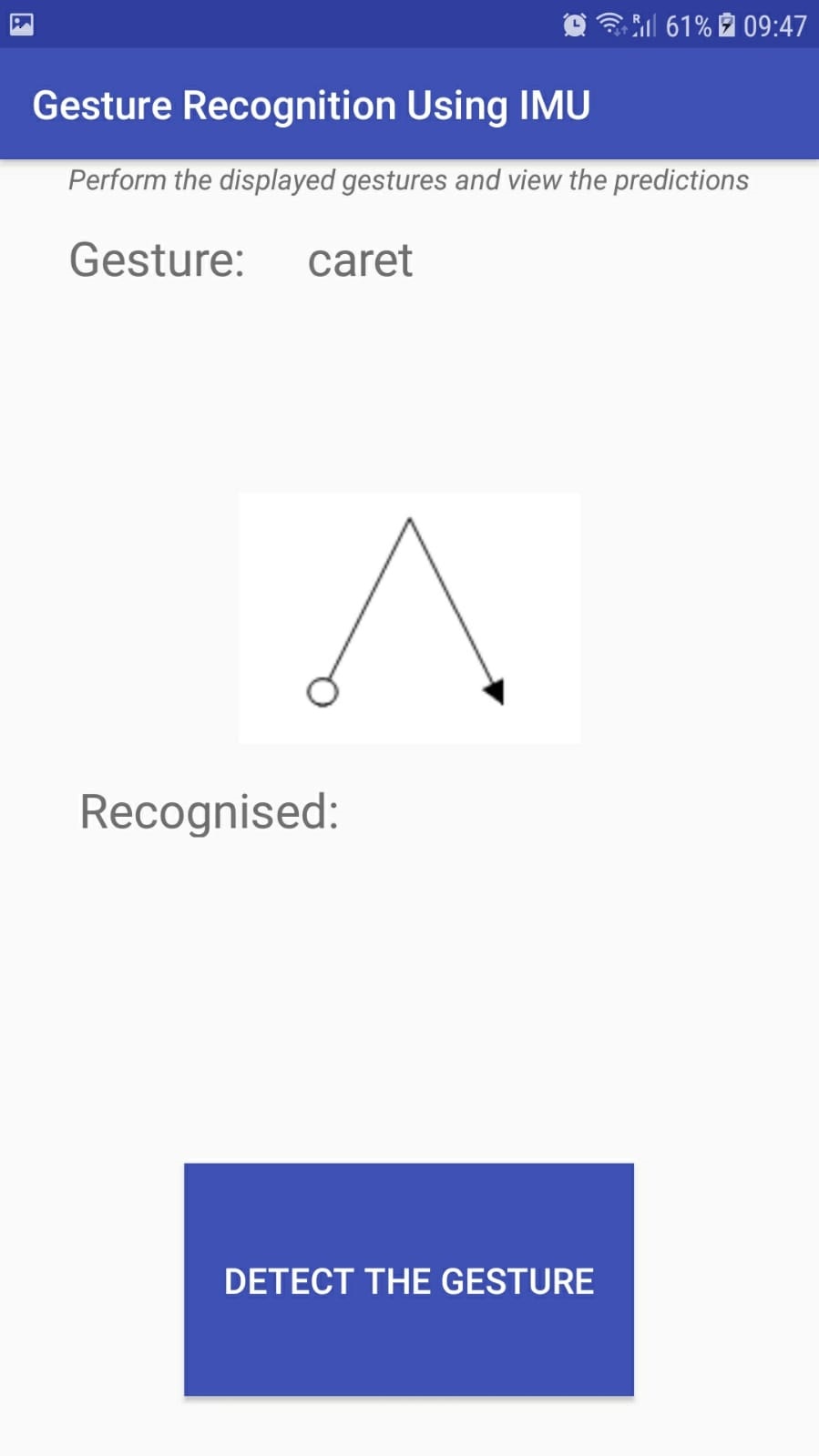}
		\caption{Gesture recognition}
		\label{fig:GestureDetect}
	\end{subfigure}
\caption{Views of various pages of the Gesture Recognition Application}
\end{figure}

\subsection{Tasks}
The User was presented with the Nexus5X phone running the Android application - Gesture Training. On the first screen, the user had to enter details such as - Name (an Alias name), gender, handedness (Left or right), age and Occupation.
Upon navigation to the next screen, a random number was generated based on which the gestures appear in a specific sequence. Each random number(pseudo-random set hard-coded in Android) corresponds to a unique gesture pattern set, which ensures that the user is not biased to a specific gesture pattern.

The supervisor, handling the phone-2, enters the user credentials and the random number on the second phone, captures the start and end of each gesture, in addition to ensuring that the user performs the right gestures. For uniformity, it was ensured that the initial orientation of the phone was parallel to the ground (the xy-plane parallel to the ground)(Fig:\ref{fig:Apparatus}).

\subsection{Time Synchronization}
The two phones had to be precisely synchronized in order to capture the exact time stamps of each gesture. In order to do that, an NTP protocol setup was experimented. System clock of Phone-1 was broadcast through the NTP server (TimeServer App\cite{TimeServer}) over WiFi Hotspot. Phone-2 connects to this server over WiFi, following which it synchronizes its System clock with the received time (with the help of Clock Sync App\cite{ClockSync}). At the beginning of each dataset (every 10 sets per User), the phones were synchronized in order to address possible differences in System clocks due to their Internal Hardware Clock tolerances.

\subsection{Participants}
The data collection process involved 19 participants in total. The participants were students who use their smartphones throughout the day for different purposes including taking pictures, scanning documents, messaging, but none of the them had used smartphone gestures for an explicit application. The 19 participants (3 female and 16 Male) were between 24 and 28 years old. Each participant provided 30 gesture sets on average. All the participants recorded the data with their right hands (irrespective of their dominant hand) for uniformity. 

\section{Data Preprocessing}
Following the data collection phase, firstly, corrupted and incorrect files were removed. Then, the timestamps from the phone-2 was used to appropriately segregate different gestures from the complete set, which resulted in 10 individual files (for 10 gestures) per dataset per participant. Each of these files were further cut into fixed-lengths corresponding to the size of the sliding window, forming a 3-Dimensional array (Number-Of-Datasets x Window-Size x 6, where 6 corresponds to the number of axes of the sensors). The size of the sliding window was chosen to be 250 samples, with moving step size of 50 samples. Hence a gesture with an average length of 612 samples would be considered as 8 individual inputs.

\begin{figure}
	\begin{subfigure}[b]{0.40\textwidth}
		\includegraphics[width=\textwidth]{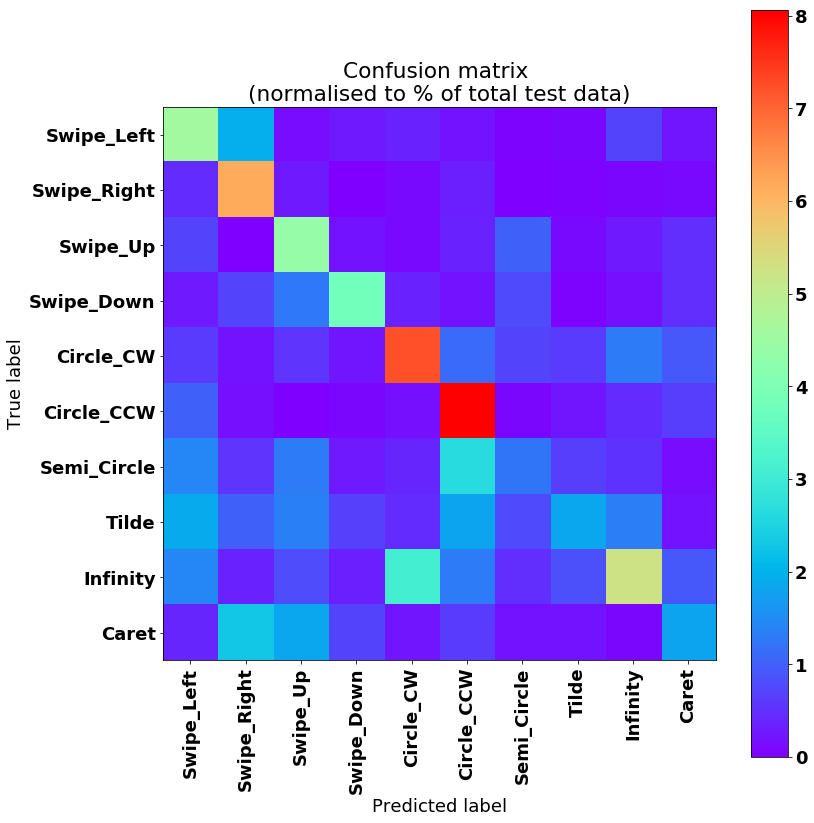}
		\caption{Confusion Matrix - Complete data set, Accuracy = 44.33\%}
		\label{fig:UserInfo}
	\end{subfigure}
	\begin{subfigure}[b]{0.40\textwidth}
		\includegraphics[width=\textwidth]{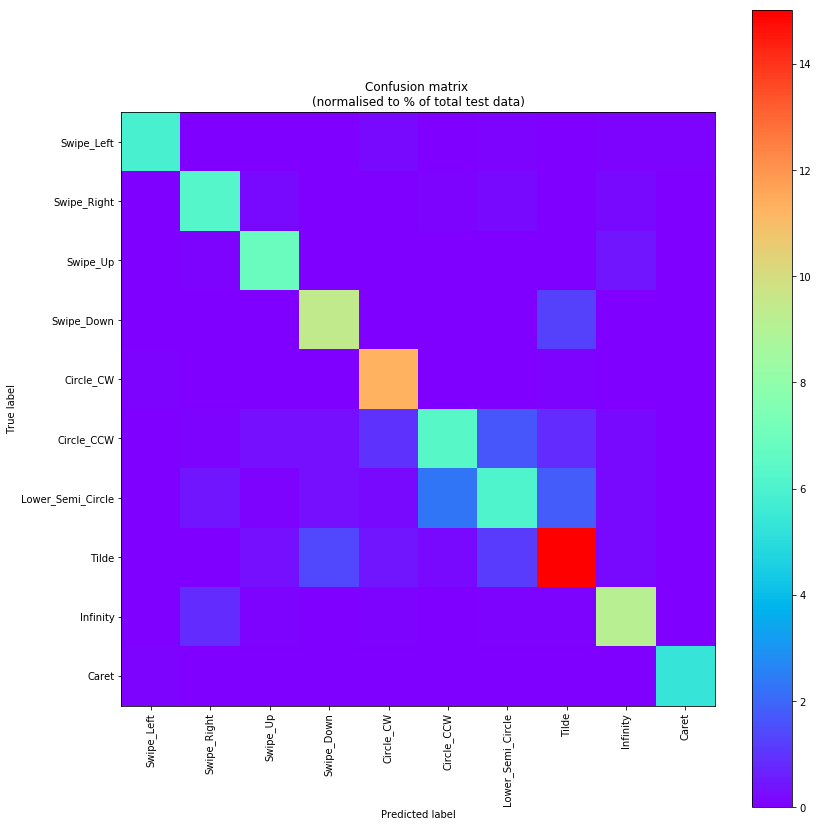}
		\caption{Confusion Matrix - Limited data set, Accuracy = 91.54\%}
		\label{fig:MainPage}
	\end{subfigure}
	\caption{Confusion Matrices for different datasets}
\end{figure}

\section{Training and Validation datasets}
Three out of the total number of participants were considered for validation, while others were used for training. The trained model showed a validation accuracy of 44.33\%. Further analysis of the collected data revealed large variances in the data. This large training data variances can only be compensated with large amount of data. 

In order to compensate the deficit in data, a limited-data-set was carefully chosen from participants with lesser variances. This included data from 3 participants for training and 1 participant for testing. Performance analysis and comparison between the models were done based on this limited-dataset.

\section{Results}
Different models were experimented on Keras\cite{Keras} and Tensorflow\cite{Tensorflow} platforms. During the initial phases of the project, in Tensorflow, a model with input ReLU activation layer and two stacked LSTMs, each with output shape of 32 were used. The model was trained using Adam optimizer with batch size of 50 and at a learning rate of 0.025. Later on, keras was chosen due to the ease of tuning and manipulation of multiple layers. In keras, a sequential model with two LSTM with 64 output neurons, followed by a dropout layer, then an LSTM layer with an output size 64 and finally a dense layer with Softmax activation was used. The Neural network was trained with Adam as the optimizer with a batch size of 50 and used the categorical cross entropy as the loss function.

\begin{marginfigure}
	\includegraphics[width=\marginparwidth]{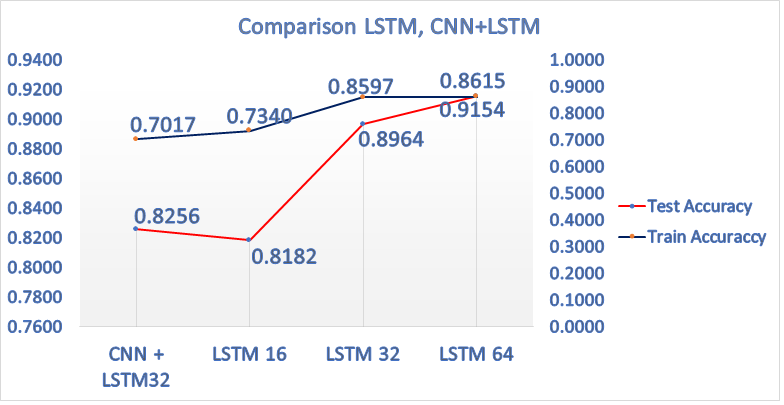}
	\caption{Comparison between models using limited-dataset}
	\label{fig:LSTMCompare}
\end{marginfigure}

When using the full data set, the model resulted in a validation accuracy of 44.33\% and a range of accuracies from 81.8 to 91.5\% using limited-dataset. The comparison between the different models used is shown in figure:\ref{fig:LSTMCompare}.

\section{Discussion}
The results show that datasets with lesser variances (limited gestures with controlled speed and rotation) leads to a better fit of the model. There are other attributes which when considered had influences on the final output.

\begin{itemize} 
	\item 	
		Selection of gestures : For gesture recognition using LSTM based on sliding window, classification is affected if one gesture is part of another. For example, Tilde is a part of Infinity and Semicircle is a part of Circle(Fig:\ref{fig:gestures02}). The sliding window would only make a segment of a gesture visible and hence leads to misclassification.
	\item 	
	Normalization : In another attempt to improve accuracy, Z-Score normalization was used to bring the data to zero-mean and unit-variance on individual axis. This step did not result in improvement in accuracy since the relative variance between data from different users still remained the same. Batch normalization layers in the keras model also did not show improvement in accuracy.
\end{itemize}

\section{Conclusion and further work}
A Sliding window based Multilayer LSTM for gesture recognition was presented as part of this project. Among other models, LSTM has showed significant results, especially for timeseries based data. Hence, most of the focus was on tuning different LSTM models and preprocessing data in order to achieve better results. 

Further work on this topic includes considering Acceleration without gravity. This would help in getting rid of the mean acceleration due to gravity on the axis that is parallel to the ground. Another approach would be to completely eliminate one of the axis, as though the gesture was made perpendicular to a plane. 

The gesture recognition in this project targets the use of readily available IMUs on an unmodified smartphone. The usecases span from controlling powerpoint presentations and home appliances to unlocking smartphones using gestures. An exhaustive dataset would further improve the accuracy and hence the confidence of the output in order to achieve expected classification of gestures in realistic applications.

%
%
%
%
%
%
%
%

\bibliography{bibliography}
\bibliographystyle{ACM-Reference-Format}

\end{document}